\documentclass[aps,superscriptaddress,twocolumn,floatfix]{revtex4}
\usepackage[centertags]{amsmath}
\usepackage{amsthm,amssymb,bbm,graphicx,subfigure,hyperref}
\pdfoutput=1

\newcommand{\one}{\mathbbm{1}}
\newcommand{\del}{\partial}
\newcommand{\nn}{\notag}
\newcommand{\braket}[1]{{\left\langle{#1}\right\rangle}}
\newcommand{\abs}[1]{{\left\vert{#1}\right\vert}}
\newcommand{\QCD}{\text{QCD}}
\newcommand{\RMT}{\text{RMT}}
\newcommand{\D}{\mathcal D}
\newcommand{\N}{\mathcal N}
\newcommand{\C}{\mathcal C}
\renewcommand{\P}{\mathcal P}
\newcommand{\T}{\mathcal T}
\newcommand{\U}{\text{U}}
\newcommand{\SU}{\text{SU}}
\renewcommand{\phi}{\varphi}
\renewcommand{\epsilon}{\varepsilon}
\DeclareMathOperator{\tr}{tr}
\begin{document}

\title{Topology and chiral random matrix theory at nonzero imaginary chemical potential}

\author{C. Lehner}
\affiliation{Institute for Theoretical Physics, University of Regensburg, 93040 Regensburg, Germany}
\author{M. Ohtani}
\affiliation{Kyorin University, School of Medicine, Mitaka, Tokyo 181-8611, Japan}
\author{J.J.M. Verbaarschot}
\affiliation{Institute for Theoretical Physics, University of Regensburg, 93040 Regensburg, Germany}
\affiliation{Department of Physics and Astronomy, SUNY, Stony Brook, New York 11794, USA}
\author{T. Wettig}
\affiliation{Institute for Theoretical Physics, University of Regensburg, 93040 Regensburg, Germany}

\date{February 16, 2009}

\begin{abstract}
We study the effect of topology for a random matrix model of QCD
 at nonzero imaginary chemical potential or nonzero temperature. 
Nonuniversal fluctuations of Dirac eigenvalues lead to normalization
factors that contribute to the $\theta$ dependence of the partition 
function. These normalization factors have to be canceled
in order to reproduce the $\theta$ dependence of the QCD
partition function. The reason for this behavior is that the 
topological domain of the Dirac spectrum (the region of the Dirac
spectrum that is sensitive to the topological charge) extends beyond
the microscopic domain at nonzero imaginary chemical
potential or temperature. 
Such behavior could persist in certain lattice formulations of
QCD.
\end{abstract}

\maketitle

\section{Introduction}

Chiral random matrix models \cite{Shuryak:1992pi,Verbaarschot:1994qf} 
have been used with great success in the past 15 years to understand and
compute universal features of the QCD Dirac spectrum; see
Ref.~\cite{Verbaarschot:2000dy} for a review.  The effect of
temperature or chemical potential can be included in these models in a
schematic way to obtain qualitative, nonuniversal results for the
QCD phase diagram.  The main purpose of this paper is to point out and
clarify a number of subtleties and nonuniversal effects that can
arise when the effects of topology become important in such schematic
random matrix models.  In particular, we shall see that nontrivial
normalization factors can occur which will be related to nonuniversal
properties of the Dirac spectrum.

Let us first consider QCD at zero temperature with $N_f$ quark
flavors, which for simplicity we take to be of equal mass $m$.  The
QCD partition function, $Z^\QCD$, can be considered at fixed
$\theta$ angle or at fixed topological charge $\nu$.  In the former
case, the $\theta$ angle can be introduced according to (see, e.g.,
\cite{Leutwyler:1992yt,Verbaarschot:2000dy})
\begin{align}
  m_R \to m \, e^{i\theta/N_f}\,, \quad m_L \to m \, e^{-i\theta/N_f}\,,
  \label{eqn:theta-gen}         
\end{align}
where $m_R$ ($m_L$) is the mass that couples right-handed (left-handed) 
quarks with antiquarks of opposite chirality.
We assume $m$ to be real and positive.  

If the number of right-handed and left-handed modes differs by $\nu$,
the product of the fermion determinants results in an overall factor
$e^{i \nu \theta}$, and we have
\begin{align}
  Z^\QCD(m,\theta)=\sum_{\nu=-\infty}^\infty e^{i\nu\theta}Z_\nu^\QCD(m)\,.
  \label{eqn:ZQCDtheta}
\end{align}
This relation can be inverted to give the QCD partition function at
fixed $\nu$,
\begin{align}
  Z^\QCD_\nu(m) = \frac 1{2\pi} \int_0^{2\pi} d\theta \: e^{-i\nu \theta}
  Z^\QCD(m,\theta)\,,
\end{align}
which corresponds to a path integral restricted to gauge fields of
topological charge $\nu$.

It is generally assumed that the gauge field measure does not depend
on the topological charge. When topological excitations can be
considered as independent events, the central limit theorem dictates
that the distribution of topological charge is given by
\begin{align}
{\cal P}_\nu = \frac 1{\sqrt{2\pi V \chi}} e^{- {\nu^2 }/{2 V \chi}}\,,
  \label{eqn:Pnu}
\end{align}
where $V$ is the space-time volume and $\chi$ is the
topological susceptibility at $\theta=0$.   
In the quenched theory (or, equivalently,
for heavy quarks), $\chi=\chi_q$ is a mass-independent constant,
whereas for light quarks, the topological charge is screened, 
resulting in a topological susceptibility at $\theta = 0$ given
by \cite{Crewther:1977}
\begin{align}
  \label{eqn:chim}
  \chi = m \Sigma\,,  
\end{align}
where $\Sigma$ is the absolute value of the 
chiral condensate for $m=0$ and $\theta=0$. 

In the microscopic  domain of QCD, where the
Compton wavelength of the pion is much larger than the size of the
box, the mass and $\theta$ dependence of the QCD
partition function is given by a random matrix theory (RMT) with the
same global symmetries as those of QCD.  Contrary to QCD, random
matrix partition functions are defined in terms of integrals over the
matrix elements of the Dirac operator at {\em fixed\/} topological
charge rather than integrals over gauge fields at fixed
$\theta$ angle, which contain the sum over topological charges.  
In this paper we will study random matrix theories that are deformed
by an imaginary chemical potential or temperature. The deformation
parameter will be denoted by $u$.

Given
a random matrix partition function at fixed $\nu$, the partition
function at fixed $\theta$ is 
defined by
\begin{align}
  Z^\RMT(m, \theta) = \sum_{\nu=-\infty}^\infty e^{i\nu \theta} 
  \P_\nu \N_\nu Z^\RMT_\nu(m)\,,
  \label{eqn:part-theta}
\end{align}
where we separated a potentially nontrivial normalization factor
$\N_\nu $ and 
a weight factor
$\P_\nu$ from $Z^\RMT_\nu(m)$.
The factor $\P_\nu$ corresponds to the
quenched distribution of topological charge given in
Eq.~\eqref{eqn:Pnu} (with $\chi=\chi_q$).  
Other $\nu$-dependent normalization
factors that may arise
in random matrix models of the QCD partition function are included
in the factor $\N_\nu$. 
One of
our objectives is to discuss the significance of these two factors.  
We shall see in
Sec.~\ref{sec:solA} that, contrary to QCD or 
chiral random matrix theories at $u = 0$
\cite{Shuryak:1992pi,Damgaard:1999ij},
$\N_\nu$ can become a nontrivial 
function of the deformation parameter. 
On the other hand, 
as  will be shown in Sec.~\ref{sec:pnu}, for light quarks it makes no
difference whether or not $\P_\nu$ is included 
in the sum over $\nu$ \cite{Damgaard:1999ij}.

A related question we would like to address in this paper is which part 
of the Dirac spectrum is sensitive to the topological charge.  
The answer to this question could depend on the parameters of
QCD or the chiral random matrix model, 
and we shall see below that this
is actually the case.  It also depends on the value of the quark mass,
for which we distinguish the following scales: (i) The microscopic scale
\cite{Shuryak:1992pi,Verbaarschot:1994qf}
where $mV \Sigma$ is kept fixed in the
thermodynamic limit.  This corresponds to 
the $\epsilon$ regime of
chiral perturbation theory \cite{Gasser:1987ah}.  
(ii) The chiral scale where $m\sqrt V$ is kept fixed in the thermodynamic
limit. This corresponds to the $p$ regime
\cite{Gasser:1986vb} of chiral perturbation theory. 
(iii) The macroscopic domain with $m \sim
\Lambda_\text{QCD}$. 
In the microscopic domain, the mass dependence of the QCD partition function is
given by chiral random matrix theory. Actually, this domain extends beyond
the microscopic domain all the way to the chiral scale. Therefore, it is
appropriate to borrow the name ``ergodic domain''
from the theory of disordered systems  \cite{Guhr:1997ve} to  
distinguish
the domain  $m \ll 1/\Lambda_{\rm QCD} \sqrt V$ from the microscopic scaling
domain. 
Note that we will sometimes consider the limit where
$mV\Sigma$ approaches infinity 
with the understanding that the thermodynamic limit
is taken first so that 
$m$ is still in the
microscopic domain.

The issues that will be addressed in this paper are already manifest
for one quark flavor, and for simplicity we will only discuss this
case. The one-flavor QCD partition function, given by the average
fermion determinant, is a function of the quark mass and of the
$\theta$ angle or the topological charge $\nu$.  If the eigenvalues of
the (anti-Hermitian) Dirac operator at fixed $\nu$ are denoted by
$i\lambda_k^{\nu}$, the QCD partition function at fixed $\nu$ can be
expressed as
\begin{align}
  \label{eqn:zev}
  Z_\nu (m)=\biggl\langle \prod_k (i\lambda_k^{\nu} + m) \biggr\rangle\,,
\end{align}
where the average is over gauge fields with fixed $\nu$.

We know that in the microscopic domain (and in fact in the ergodic domain) 
the mass dependence of the               
one-flavor QCD partition function in the sector of topological charge
$\nu$ is given by \cite{Gasser:1987ah,Leutwyler:1992yt}
\begin{align}
  \label{eqn:univznu}
  Z_\nu(m) \sim I_\nu(mV\Sigma) \,.
\end{align}
For large values of the argument the modified Bessel function $I_\nu$
becomes insensitive to its index $\nu$, and thus
Eq.~\eqref{eqn:univznu} implies that the average fermion determinant
does not depend on the topological charge when $mV\Sigma\gg1$.  
In terms of Dirac eigenvalues one way to realize this is 
when only eigenvalues below this mass scale
are affected by topology [see Eq.~\eqref{eqn:zev}].
However, more exotic scenarios
are also possible. It could be that eigenvalues beyond the microscopic
domain are sensitive to the topological charge. If $m$ is in the microscopic 
domain, this might result in a $\nu$-dependent overall factor 
$\N_\nu$ that could 
depend on the 
deformation parameter $u$
and restores the $\nu$-independence of $Z_\nu$ for $mV\Sigma \gg 1$.
To find out whether this  scenario is realized, it
makes sense to introduce the notion of the {\em topological domain of
  the Dirac spectrum}, which we define to be the part of the Dirac
spectrum that is sensitive to the topological charge.

In QCD we have $\N_\nu=1$ and, from Eqs.~\eqref{eqn:ZQCDtheta} and
\eqref{eqn:univznu}, 
the universal
$\theta$ dependence of the partition function is given by
\begin{align}
  \label{eqn:ZQCDthetauniv}
  Z^\QCD(m,\theta)\sim e^{mV\Sigma\cos\theta}\,.
\end{align}
It is plausible that the standard scenario discussed after
Eq.~\eqref{eqn:univznu} applies in this case, i.e., the topological
domain of the Dirac spectrum does not extend beyond the microscopic
domain. Exotic scenarios such as the one discussed above could occur
in certain lattice formulations of QCD, 
and it would be interesting to test this directly.  We
shall further comment on this point in the conclusions.

The ergodic domain of QCD is given by random matrix theory,
but since the average fermion determinant is sensitive to {\em all\/}
eigenvalues, it could be that deformations of the random matrix model
result in a topological domain that extends beyond the microscopic
domain.  In this paper we will see that this may happen in random
matrix models at nonzero temperature/imaginary chemical potential.
 
The $\theta$ dependence of random matrix theories at nonzero
temperature was discussed before in the literature
\cite{Janik:1999ps}. In that work the temperature was introduced such
that it only affects the eigenmodes corresponding to nonzero Dirac
eigenvalues.  This resulted in the same $\theta$ dependence as in the
zero-temperature random matrix model \cite{Shuryak:1992pi}.  Among
others it was shown that the $O(m^2)$ term in the chiral Ward
identity does not contribute in the chiral limit. This is not always
the case.  It was recently shown in the framework of chiral
perturbation theory that in the superfluid phase of QCD at nonzero
chemical potentials the
$O(m^2)$ term cannot be neglected \cite{Metlitski:2005}. In this paper
we will see that the $O(m^2)$ term in the chiral Ward identity contributes
to the topological susceptibility 
for random matrix partition functions at nonzero
temperature/imaginary chemical potential if the $u$-dependent normalization
factor $\N_\nu$ is not included.

The structure of this paper is as follows.
Chiral random matrix theories at zero and nonzero deformation parameter
will be introduced in Sec.~\ref{sec:rmt}. The random matrix models are solved
in Sec.~\ref{sec:rmtsol}, where we also discuss the normalization factor 
$\N_\nu$ and the distribution of the topological charge 
$\P_\nu$. In Sec.~\ref{sec:condtop} we show that the chiral condensate for
one flavor only has the correct behavior if the normalization
factor $\N_\nu$ is included. The origin of $\N_\nu$ is
studied in Secs.~\ref{sec:ev} and \ref{sec:susc}. In Sec.~\ref{sec:ev}
we show that it is related to the extent of the topological domain, and 
in Sec.~\ref{sec:susc} we find that the contribution of the pseudoscalar susceptibility 
does not vanish if $\N_\nu$ is not included. 
Concluding remarks are made in Sec.~\ref{sec:conc}.

\section{Chiral random matrix models}
\label{sec:rmt}

\subsection{Definition of the random matrix model}

The random matrix model for $N_f=1$ in the sector of topological
charge $\nu$ is defined by \cite{Shuryak:1992pi}
\begin{align}
  Z_\nu(m) = \C_{N, \nu}\int \D W \det (D+m)\, e^{-(1/2) N\Sigma^2 \tr W^\dagger W}
  \label{eqn:zrmt}
\end{align}
with the random matrix Dirac operator defined by 
\begin{align}
  D = \begin{pmatrix} 0 & iW \\ iW^\dagger & 0 \end{pmatrix}.
  \label{eqn:dirac}
\end{align}
The integral $\D W$ is over the real and imaginary parts of the
elements of the random matrix $W$, which has dimension $p\times q$.  The
Dirac operator \eqref{eqn:dirac} has $\abs{p-q}$ exact zero modes.  For
this reason we interpret
\begin{align}
  \nu = p- q
  \label{eqn:enu}
\end{align}
as the topological charge. The total number of modes
\begin{align}
  N= p+q
  \label{eqn:eN}
\end{align}
will be interpreted as the volume. This corresponds to the choice of 
mode density 
\begin{align}
  \frac NV = 1\,.
\end{align}
The normalization factor $\C_{N, \nu}$ is chosen such that the quenched partition
function  is normalized to unity, i.e.,
\begin{align}
\C_{N, \nu} = \left(\frac {N\Sigma^2}{2\pi}\right)^{(1/4)(N^2-\nu^2)}\,.
\end{align}

We will consider this random matrix model in the presence of an
imaginary chemical potential $iu$.  Using the chiral representation of
the $\gamma$ matrices, the $u$-deformed Dirac operator is given by
\cite{Jackson:1995nf,Wettig:1995-1996,Stephanov:1996he,Stephanov:1996ki}
\begin{align}
  D(u) &= \begin{pmatrix}
     0 & iW + i u\, \one_{p\times q}
    \\ iW^\dagger + i u\, \one_{q\times p} & 0
  \end{pmatrix},
  \label{eqn:diracu}
\end{align}
where $\left(\one_{p\times q}\right)_{k\ell}=\delta_{k\ell}$.
Alternatively, $u$ can be interpreted as a schematic temperature as
was done in \cite{Jackson:1995nf,Wettig:1995-1996,Stephanov:1996he}. The argument goes
as follows.  The temperature enters in the Dirac operator through the
matrix elements corresponding to $\del_0$, with eigenvalues that are
given by the Matsubara frequencies.  We include only the temperature
dependence given by the lowest two Matsubara frequencies by adding the
$p\times q$ temperature matrix $i\T$ to $iW$ and $iW^\dagger$ in
Eq.~\eqref{eqn:dirac}, where
\begin{align}
  \T_{kk}  &= \left\{
  \begin{array}{cl}
    u & \text{ for } k \le \min\{p, q\}/2\,, \\
    -u & \text{ for } k > \min\{p, q\}/2\,,
  \end{array}\right.
\end{align}
and $\T_{k\ell} = 0$ for $k \ne \ell$.  Using the invariance of the
integration measure under unitary transformations $ W \to U W V^{-1}$
with $U \in \U(p)$ and $V \in \U(q)$, the temperature matrix can be
transformed into a diagonal matrix with all diagonal matrix elements
equal to $u$, so that the Dirac operator is given by
Eq.~\eqref{eqn:diracu}.

In the following, we shall refer to the model defined by
Eq.~\eqref{eqn:diracu} as model A.

\subsection{Other random matrix models}

Equation~\eqref{eqn:diracu} is not the only way to introduce a nonzero
temperature. Another possibility \cite{Janik:1999ps} is to first
partition the $N$ modes into $N_0 = p + q$ ``zero'' modes and a
fixed number $N_1$ of ``nonzero'' modes, with $\abs{\nu} = \abs{p - q}$ actual
zero modes of the Dirac operator.  An $N_1\times N_1$ temperature
matrix is then added to the nonzero-mode component of the Dirac
operator, while the zero-mode matrix elements remain temperature
independent.  In terms of the Dirac operator \eqref{eqn:diracu} this
means that we add to an $(N_1/2+p)\times(N_1/2+q)$ random matrix $W$ a
diagonal matrix with $N_1/2$ elements equal to $iu$ and $\min\{p,q\}$
elements equal to zero.  (This is technically equivalent to the model
considered in Ref.~\cite{Wettig:1995-1996}, although the physics
background is different.)  In the following, we shall refer to this
model as model B.

A third possibility is to add to $W$ a random matrix with matrix
elements that are proportional to $u$. This model was introduced in
Ref.~\cite{Osborn:2004rf} for imaginary $u$ (i.e., real chemical
potential) to describe the microscopic domain of QCD at nonzero baryon
chemical potential.  For real $u$, this results in a model that
differs from the original model \eqref{eqn:dirac} simply by a
rescaling of the parameter $\Sigma$ according to $\Sigma \to
\Sigma/{\sqrt{1+u^2}}$.  This model will be referred to as model C.
Note that this model does not have a chiral phase transition.  A less
trivial model is obtained by introducing two or more different
imaginary chemical potentials \cite{Akemann:2006ru}, but we will not
discuss this possibility in this paper.

\section{Solution of the Random Matrix Models and Normalization Factors}
\label{sec:rmtsol}

In this section we solve the random matrix models that were introduced in the
previous section.  We will find that the universal $\theta$ dependence
is not recovered for model A at $u \ne 0 $ unless additional normalization 
factors are included.

\subsection{Solution of model A}
\label{sec:solA}

In this subsection we solve the random matrix model A given by
Eq.~\eqref{eqn:zrmt} with Dirac operator \eqref{eqn:diracu}.  The
procedure is standard (see, e.g.,
\cite{Shuryak:1992pi,Jackson:1995nf}).  We start by writing the
determinant as a Grassmann integral,
\begin{align}\nn
&\det(D(u) + m)\\
&\quad =\int d\psi d \bar \psi
 \:\exp \left[
\begin{pmatrix} \bar \psi_L \\ \bar\psi_R \end{pmatrix}^T
(D(u) + m)
\begin{pmatrix} \psi_R \\ \psi_L \end{pmatrix}\right],
\end{align}
and perform the Gaussian
average over the random matrix elements.  After a Hubbard-Stratonovich
transformation and integration over the Grassmann variables we obtain
the following $\sigma$ model:
\begin{align}
  Z^A_\nu(m) &= \int  d\sigma d\sigma^* \left ( 1 + u^2 \abs{\sigma+m}^{-2}
  \right )^{n} \nn\\&\quad \times
  (\sigma+m)^p (\sigma^*+m)^q \, 
  e^{-(1/2)N\Sigma^2 \sigma \sigma^*}\,,
  \label{eqn:zsig}
\end{align}
where $n = \min\{ p, q \}$.  Notice that the 
$\nu$-dependent normalization constant
introduced in Eq.~(\ref{eqn:zrmt}) has canceled.

After changing variables $\sigma \to \sigma - m$ and $ \sigma^* \to
\sigma^* - m$ in Eq.~\eqref{eqn:zsig} and then expressing the integral
over $(\sigma,\sigma^*)$ in polar coordinates $(r,\phi)$, the angular
integral results in a modified Bessel function, and the partition
function is given by the remaining integral over $r$,
\begin{align} \nn 
  Z^A_\nu(m) = 2\pi \int_0 ^\infty & dr\:
  I_\nu(mN\Sigma^2r)
  r^{\abs{\nu}+1} (r^2+u^2)^{(N-\abs{\nu})/2} \\
  & \times e^{-(1/2) N\Sigma^2(r^2 + m^2)}\,.
  \label{eqn:znurad}
\end{align}
For large $N$, this partition function can be evaluated by a
saddle-point approximation.  For $m$ in the ergodic domain, the
saddle point in the broken phase is at $\bar r^2 = 1/\Sigma^2 -u^2$.
To leading order in $1/N$ the partition function is given by 
\begin{align}
  Z^{A,\text{as}}_\nu(m) \sim
  I_\nu\big(mN\Sigma^A(u)\big) \tau^\abs{\nu}\,,
  \label{eqn:znusad}
\end{align}
where irrelevant prefactors have been ignored and 
\begin{align}
  \Sigma^A(u)=\Sigma\tau(u)\quad\text{with}\quad
  \tau(u) = \sqrt{1-\Sigma^2 u^2}\,.
\end{align}
A second-order phase transition to the chirally symmetric phase occurs
at $u_c=1/\Sigma$ \cite{Jackson:1995nf}.

The $\theta$ dependence of the partition function is obtained after performing
the sum over $\nu$ according to Eq.~\eqref{eqn:part-theta}. 
As will be explained in detail in the next subsection, 
for light quarks the sum is not
affected by the distribution function $\P_\nu$ \cite{Damgaard:1999ij}.
We will therefore set $\P_\nu=1$.

Let us first consider the case $u =0$ and take $\N_\nu = 1$.  Using
the identity for Bessel functions given by
(\cite{abramowitz+stegun}, Eq.~(9.6.33))
\begin{align}
  \label{eqn:sum1}
  \sum_{\nu=-\infty}^\infty I_\nu(x) \: t^\nu = e^{(1/2) x(t+1/t)}\,,
\end{align}
we find the universal result \cite{Leutwyler:1992yt,Damgaard:1999ij}
\begin{align}
  Z^A(m,\theta)\big|_{u=0} \sim e^{mN \Sigma \cos \theta}\,.
  \label{eqn:Zthetauniv}
\end{align}
This shows that we do not need nontrivial normalization factors at $u=0$.

Now consider the case $u\ne0$.
Because of the factor $\tau^{\abs{\nu}}$, in this case
Eq.~\eqref{eqn:znusad} depends on $\nu$ for
$mN\Sigma^A(u)\gg1$.   This is a
nonuniversal result and would also lead to a nonuniversal
$\theta$ dependence of $Z^A$ after summing over $\nu$.  However, these
problems can be fixed by introducing a $u$-dependent normalization
factor
\begin{align}
  \N_\nu = \tau^{-\abs{\nu}}\,.
  \label{eqn:znor}
\end{align}
Then with the replacement $\Sigma \to \Sigma^A(u)$  the sum over 
$\nu$ is the same as for $u = 0$. Again the sum is not affected by
the distribution function $\P_\nu$, and we find the universal result
\begin{align}
  Z^A(m,\theta) \sim e^{mN \Sigma^A(u) \cos \theta}  \,.
\end{align}
In QCD an imaginary chemical potential is equivalent to a constant vector
field and can be gauged into the temporal boundary conditions of the 
fermion fields. 
This is not the case in random matrix theory, and therefore it should
not come
as a surprise that we need a $\nu$-dependent normalization factor to
recover the correct $\theta$ dependence. In agreement with
universality properties of Dirac spectra at fixed $\nu$
\cite{Jackson:1996xt,Jackson:1997ud,Guhr:1997uf, Akemann:2006ru} this
normalization factor does not depend on the quark mass.

When $u$ approaches $u_c = 1 / \Sigma$, higher-order terms in the saddle-point
approximation of Eq.~\eqref{eqn:znurad} become important, and the integral has
to be performed exactly. We will not further elaborate on this and 
only discuss the parameter domain where the leading-order saddle-point
approximation is appropriate.

We will discuss further properties of model A in later sections but
first turn to a discussion of the necessity of $\P_\nu$ and to
a comparison with models B and C, where no $u$-dependent normalization
factors will be needed.

\subsection{On the necessity of $\P_\nu$}
\label{sec:pnu}

For large $\abs{\nu}$ at fixed $x$ the modified Bessel function can 
be approximated by (\cite{abramowitz+stegun}, Eq.~(9.3.1))
\begin{align}
I_\nu(x) \sim \frac {(x/2)^{\abs{\nu}}}{\abs{\nu}!}\,.
\end{align}
Therefore, if $m$ is in the microscopic domain, the sum over $\nu$ in
Eq.~\eqref{eqn:part-theta} is convergent without the Gaussian
factor \eqref{eqn:Pnu}.  

The sum over $\nu$ can be performed, up to exponentially suppressed
contributions, using the approximation \cite{Leutwyler:1992yt} 
\begin{align}
\label{eqn:ias}
I_\nu(x) \sim \frac 1{\sqrt{2\pi x}}\, e^{x - \nu^2/2x}\,,
\end{align}
which follows from the uniform large-order expansion of the modified
Bessel function and is valid for $ 1 \ll \abs{\nu} \ll x$
(\cite{abramowitz+stegun}, Eq.~(9.7.7)).  It makes no difference whether
or not we include the factor $\P_\nu$ in Eq.~\eqref{eqn:part-theta}
since
\begin{align}
  e^{-(\nu^2/2N)\left((1/m\Sigma(u))+(1/\chi_q)\right)}
  \sim  e^{-(\nu^2/2mN\Sigma(u))}
\end{align}
for $m$ in the ergodic domain.  The topological susceptibility 
at $\theta=0$ is
therefore given by Eq.~\eqref{eqn:chim}.   From the approximation
\eqref{eqn:ias} we also see that all topological sectors with $\nu^2 \ll
mN\Sigma(u)$ contribute equally to the partition function.

It was argued by Damgaard \cite{Damgaard:1999ij} that the factor
$\P_\nu$ should be absent in the sum over $\nu$ in
Eq.~\eqref{eqn:part-theta}, although he also pointed out that the
quenched limit could not be taken properly in this case.  Our point of
view is that the presence of $\P_\nu$ is immaterial for $m$ in the
microscopic domain, but that $\P_\nu$ becomes important at length scales
below the inverse $\eta'$ mass where it is believed to determine the local
topological susceptibility and leads to the Witten-Veneziano formula
for the $\eta'$ mass 
\cite{Witten:1979vv,Veneziano:1979ec,Shuryak:1994rr,Ilgenfritz:2008private}.
Beyond this scale the topological susceptibility 
at $\theta=0$ is given by Eq.~\eqref{eqn:chim}.

\subsection{Comparison with models B and C}
\label{sec:solBC}

For fixed topological charge $\nu$ the partition function of model B
is given by
\begin{align}\nn
  Z^B_\nu(m) = \int & d\sigma d\sigma^* (\abs{\sigma+m}^2+u^2)^{N_1/2} \\
  & \times (\sigma+m)^p (\sigma^*+m)^q e^{-(1/2) N \Sigma^2
    \sigma \sigma^*}\,,
  \label{eqn:znujanik}
\end{align}
or, after introducing polar coordinates,
\begin{align}\nn
  Z^B_\nu(m) = 2 \pi \int_0^\infty & dr\: I_\nu(m N \Sigma^2 r)  r^{N_0+1}
  \left(r^2+u^2\right)^{N_1/2} \\
  &\times e^{-(1/2) N \Sigma^2 (r^2 + m^2)}\,.
\end{align}
Note that this partition function
becomes independent of $\nu$ for large $m N \Sigma$. Since 
the correct $\theta$ dependence is obtained at $u =0$ this model does
not require additional normalization factors.
The sum over
$\nu$ with $\P_\nu=1$ results in
\begin{align}\nn
  Z^B(m,\theta) = 2 \pi \int_0^\infty & dr\: e^{m N \Sigma^2 r\cos\theta}  r^{N_0+1}
  \left(r^2+u^2\right)^{N_1/2} \\
  &\times e^{-(1/2) N \Sigma^2 (r^2 + m^2)}\,.
\end{align}
 Using a saddle-point approximation for
large $N$, we find the universal $\theta$ dependence
\begin{align}
  Z^B(m,\theta) \sim e^{m N \Sigma^B(u)\cos \theta}\,,
\end{align}
where \cite{Wettig:1995-1996}  
\begin{align}
\frac{\Sigma^B(u)}{\Sigma}=\left [\frac{
 1\!-\!\Sigma^2 u^2 \!+\!\sqrt{(1\!+\!\Sigma^2u^2)^2 \!-\!4 
   \Sigma^2 u^2N_1/N}}2 \right]^{1/2}. 
\end{align}

The partition function of model C at deformation parameter $u$ is
equivalent to Eq.~\eqref{eqn:zsig} at $u=0$ with $\Sigma \to \Sigma^C(u) =\Sigma /
\sqrt{1+u^2}$, and we thus have the universal result
\begin{align}
  Z^C(m,\theta) \sim e^{m N \Sigma^C(u) \cos \theta}.
\end{align}

Hence we see that in order to obtain the universal $\theta$ dependence of 
the partition function neither model B nor model C requires normalization
factors $\N_\nu$ that depend on the deformation parameter $u$.
However, let us emphasize again that the appearance of the $\N_\nu$ is
a generic feature in RMT.  
In the remainder of this paper we will identify 
mechanisms that are responsible for this behavior.

\section{Chiral Condensate and Topology}
\label{sec:condtop}

 The case 
$N_f =1$ we address in this paper is
special since there is no $\SU(N_f)\times\SU(N_f)$ symmetry that could
be spontaneously broken.  Nevertheless, there could still be a chiral
condensate, which can be calculated in the usual way,
\begin{align}
  \label{eqn:cond}
  \abs{\braket{\bar\psi\psi}}&=\frac1V\del_m\log Z(m,\theta)\,.
\end{align}
The parameter $\Sigma$ introduced earlier is defined to be equal to
$\abs{\braket{\bar\psi\psi}}$ at $\theta=0$ for $m\to0$ and
$V\to\infty$.  The functions $\Sigma(u)$ computed in
Secs.~\ref{sec:solA} and \ref{sec:solBC} correspond to the
$u$-dependent chiral condensate in the same limits.  These limits can
be taken in different orders \cite{Shifrin:2005cy}, either
\begin{align}
  \label{eqn:sigma1}
  \Sigma^{(1)} = \lim_{V\to \infty} \lim_{m \to 0} \frac 1V \del_m
  \log Z(m,\theta=0)
\end{align}
or in the reverse order
\begin{align}
  \label{eqn:sigma2}
  \Sigma^{(2)} = \lim_{m \to 0} \lim_{V\to \infty} \frac 1V \del_m
  \log Z(m,\theta=0)\,.  
\end{align}
In Eq.~\eqref{eqn:sigma1}, a nonzero chiral condensate implies the
breaking of the $\U_A(1)$ symmetry by instantons or the chiral anomaly
\cite{'tHooft:1986nc}, whereas in Eq.~\eqref{eqn:sigma2} a nonzero
chiral condensate implies ``spontaneous symmetry breaking'' in the
following sense.  At {\em fixed\/} topology the QCD partition function
has a $\U_A(1)$ symmetry (in fact a covariance except at $\nu =0$
where we have a symmetry).  A nonzero chiral
condensate spontaneously breaks this $\U_A(1)$ symmetry at fixed
topology.  

From the universal expression (\ref{eqn:ZQCDthetauniv})
for the one-flavor partition function
it is clear that the order of limits should not matter. We will now see
that for model A this is only the case if the normalization factors
$\N_\nu$ are included. Because
in this section we only consider model A we omit the
superscript $A$.  Using Eq.~\eqref{eqn:part-theta} and the mass
dependence of $Z_\nu(m)$ given by Eq.~\eqref{eqn:znusad}, we find that
$\Sigma^{(1)}$ of model A is given by
\begin{align}
  \Sigma^{(1)} = \lim_{N\to\infty}\lim_{m\to0}
  \frac{\del_m [ \N_1 Z_{1}(m)+\N_{-1} Z_{-1}(m)]}
  {N\N_0Z_{0}(m)}\,,
\end{align}
where the factor $\P_\nu$ has dropped out of numerator and denominator
since it is essentially constant for $\nu \ll \sqrt N$.  Using the
result \eqref{eqn:znusad}, we obtain
\begin{align}
  \Sigma^{(1)}(u) = (\N_1/\N_0) \Sigma \tau^2 = (\N_1/\N_0) \Sigma (1
  - \Sigma^2 u^2)\,.
\end{align}

Next we calculate the chiral condensate using the reverse order of
limits.
Based on the discussion in Sec.~\ref{sec:pnu} we find that for
$\abs{\nu} \ll \sqrt{mN\Sigma}$ the condensate for fixed $\nu$
does not depend on $\nu$.  Its value is therefore equal to the value
in the $\nu=0$ sector. This was calculated in
Ref.~\cite{Jackson:1995nf}, resulting in
\begin{align}
  \Sigma^{(2)}(u) = \Sigma \tau = \Sigma \sqrt{ 1- \Sigma^2 u^2}\,.
\end{align}

We thus see that the two condensates are only equal
if the normalization factor
$\N_1/\N_0=1/\sqrt{1-\Sigma^2u^2}$ from Eq.~\eqref{eqn:znor} is included.  
Therefore we have
a nice consistency check of Eq.~\eqref{eqn:znor}.

So far, we have assumed that we can choose $p$ and $q$ to have
arbitrary $\nu=p-q$.  Let us now fix the total number of states $N$.
In this case the Dirac operator with $\nu$ zero modes has nonzero
off-diagonal blocks of dimension $(N+\nu)/2 \times (N-\nu)/2$; see
Eqs.~\eqref{eqn:enu} and \eqref{eqn:eN}.  This implies that the parity of
the topology is the same as the parity of $N$.  In the following we
assume that $N$, and therefore also $\nu$, is even.  
Equation~\eqref{eqn:ZQCDthetauniv} shows that 
the chiral condensate can be extracted from
\begin{align}
  (\Sigma^{(1)})^2 = \lim_{N\to \infty} \lim_{m\to 0} \frac 1{ N^2}
  \frac{\del_m^2  Z(m,\theta=0) }{Z(m,\theta=0)}\,. 
\end{align}
For $m \to 0$, the numerator receives contributions from $\nu =0$ and
$\nu = \pm 2$, while only the $\nu = 0$ sector contributes to the
denominator. For the $\nu = 0$ contribution we find \cite{Leutwyler:1992yt}
\begin{align}\nn
  (\Sigma^{(1)})^2_{\nu = 0}&= \lim_{N\to \infty} \lim_{m\to 0} \frac 1{ N^2}
  \frac{\del_m^2  Z_0(m) }{Z_0(m)} \\
  &= \lim_{N \to \infty}\frac 2 {N^2}\braket{ \sum_{k=1}^{N/2}
    \frac 1{(\lambda_k^{\nu=0})^2} }_{\!\!N_f=1}\,, 
  \label{eqn:sig20}
\end{align}
where the average includes the fermion determinant.  The right-hand side of
Eq.~\eqref{eqn:sig20} is dominated by the smallest eigenvalues.  Note
that this contribution is independent of the normalization of the
partition function.  The contributions of $\nu=\pm2$ to the condensate
are the same and can be written in terms of the Dirac eigenvalues as
\begin{align}
  (\Sigma^{(1)})^2_{\nu = \pm 2} & = 
  \lim_{N\to\infty} \frac 2{N^2} \frac {\N_{2}}{\N_0} 
  \frac{\braket{\prod_{k=1}^{N/2-1}(\lambda_k^{\nu=2})^2}}
  {\braket{\prod_{k=1}^{N/2} (\lambda_k^{\nu=0})^2}}\,,
  \label{eqn:sigt22}
\end{align}
where averages without subscript are with respect to the quenched
partition function.   
This is essentially the
ratio of the fermion determinants in the sectors $\nu=2$ and $\nu=0$.
In the random matrix model A the expressions \eqref{eqn:sig20} and
\eqref{eqn:sigt22} evaluate to
\begin{align}
  (\Sigma^{(1)})^2_{\nu = 0} & = \frac 12 \Sigma^2\tau^2\,,\\
  (\Sigma^{(1)})^2_{\nu = \pm 2} & = \frac 14 \Sigma^2 \tau^4
  \frac{\N_{2}}{\N_{0}}\,,
\label{eqn:sig22rmt}
\end{align}
so that the correct result for the chiral condensate is obtained with 
the normalization factors from Eq.~\eqref{eqn:znor}.

The question we wish to address in the sections below is why model A
requires the $u$-dependent normalization factors \eqref{eqn:znor}.  We
will relate this question to the properties of the Dirac eigenvalues.
As we have already discussed in the introduction, the requirement that
$Z_\nu(m)$ be independent of $\nu$ for $m V \Sigma \gg 1$ can
explain why a normalization factor $\N_\nu \ne 1$ is needed if the
topological domain of the Dirac spectrum extends beyond the
microscopic domain.  Equation~\eqref{eqn:sigt22} shows that the
consistency relation $\Sigma^{(1)} = \Sigma^{(2)}$ should also be
related to the properties of the Dirac eigenvalues, to which we turn
now.

\section{Eigenvalue Fluctuations and Microscopic Universality}
\label{sec:ev}

\begin{figure}[!t]
  \begin{center}
    \includegraphics[width=8cm]{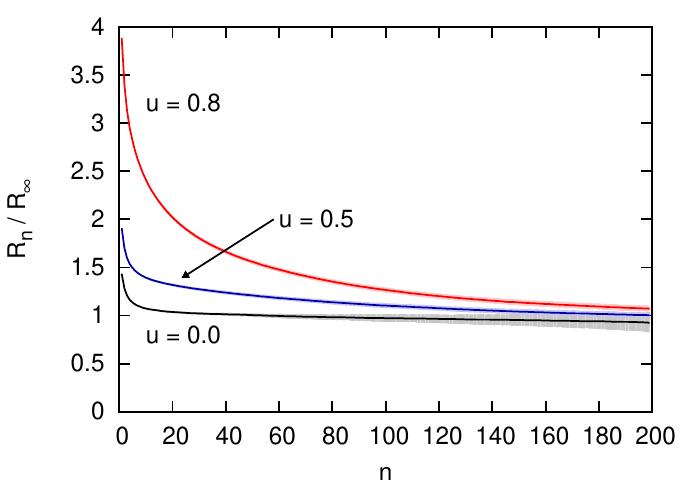}
  \end{center}
  \caption{Convergence of the ratio $R_n$ of determinants for $\nu =2$
    and $\nu = 0$ as a function of the number $n$ of eigenvalues
    included for an ensemble of $10^6$ $400\times 400$ matrices.
    Results are shown for $u=0.0$, $u =0.5$, and $u=0.8$.  The shaded
    areas correspond to the statistical errors.}
  \label{fig:fig1} 
\end{figure}

In the numerical calculation of this section we keep $N$ fixed  as discussed
at the end of the previous section.
Motivated by Eq.~\eqref{eqn:sigt22}, we consider the ratio $R_n$ of
the products of eigenvalues for $\nu = 2$ and $\nu =0$ as a function of
the number of eigenvalues included in the product,
\begin{align}
  R_n \equiv \frac 1{N^2} 
  \frac{\braket{\prod_{k=1}^{n-1}(\lambda_k^{\nu=2})^2}}
  {\braket{\prod_{k=1}^{n\phantom{1}}(\lambda_k^{\nu=0})^2}}\,.  
\end{align}
For $ n = N/2$ all eigenvalues are included in the product, and for
model A the value of this ratio follows from Eqs.~\eqref{eqn:sigt22} and
\eqref{eqn:sig22rmt},
\begin{align}
  R_\infty = \lim_{N\to \infty} R_{N/2} 
  = \frac 18 \Sigma^2\tau^4\,.
\end{align}
We have evaluated the ratio $R_n$ numerically for model A, using an
ensemble of $10^6$ random matrices \eqref{eqn:diracu} of dimension
$N=400$ distributed according to the Gaussian factor in
Eq.~\eqref{eqn:zrmt}.  The mass has been set to zero. 
In Fig.~\ref{fig:fig1} we plot the
ratio $R_n/R_\infty$ versus $n$ for $u=0$, $u=0.5$, and $u =0.8$.  We
observe that for $u=0$ the ratio of determinants saturates in the
ergodic domain ($n\lesssim\sqrt N=20$).  This is not the case for
$u=0.5$ and $u = 0.8$, where {\em all\/} eigenvalues contribute to the
ratio of the two determinants.

\begin{figure}[!t]
  \begin{center}
    \includegraphics[width=8cm]{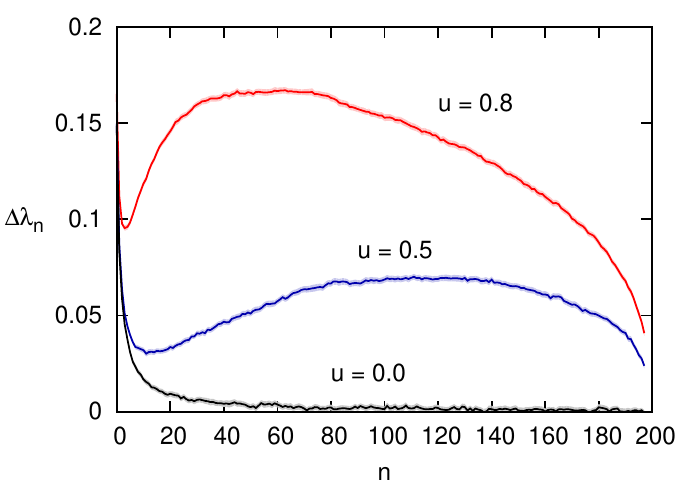}
    \includegraphics[width=8cm]{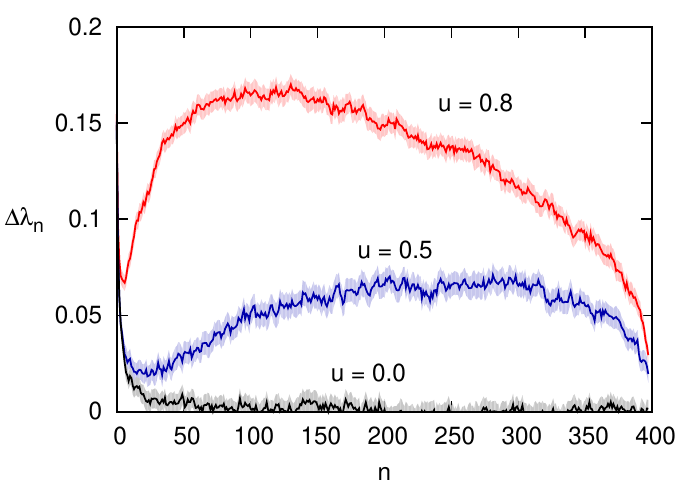}
  \end{center}
  \caption{Topological shift $\Delta \lambda_n$ of the eigenvalues for
    an ensemble of $10^6$ $400 \times 400$ matrices (top) and an
    ensemble of $10^5$ $800 \times 800$ matrices (bottom).  The shaded
    areas correspond to the statistical errors.}
  \label{fig:fig2} 
\end{figure}

This is further illustrated in Fig.~\ref{fig:fig2}, where we plot the
ratio
\begin{align}
  \Delta \lambda_n \equiv \frac {\braket{\lambda^{\nu=2}_{n}} -
    \braket{\lambda^{\nu = 0}_{n+1}}} 
  {\braket{\lambda^{\nu=0}_{n}} - \braket{\lambda^{\nu = 0}_{n+1}}}
\end{align}
versus $n$.
The motivation for constructing this particular ratio is as
follows.  The microscopic eigenvalues are expected to behave
universally after rescaling with the chiral condensate and the volume.
The universal result for the spectral density of microscopic
eigenvalues in the quenched case and in the topological sector $\nu$
is \cite{Verbaarschot:1993pm}
\begin{align}
  \rho_s(\xi) = \frac\xi2 \left[J_{\nu}(\xi)^2 
    - J_{\nu+1}(\xi)J_{\nu-1}(\xi) \right]\,,
\label{eqn:micro}
\end{align}
where  $J_\nu$ is a Bessel function and $\xi \equiv
\lambda N\Sigma$.  Its large-$\xi$ behavior is given by
\begin{align}
  \rho_s(\xi) = \frac 1 \pi - \frac{\cos(\nu\pi - 2\xi)}{2\pi\xi}
\end{align}
so that for $\lambda_n N\Sigma\gg1$ we have
$\braket{\lambda^{\nu=2}_{n}}\approx\braket{\lambda^{\nu=0}_{n+1}}$
and therefore $\Delta\lambda_n \to 0$.  This is indeed what we find in
Fig.~\ref{fig:fig2} for $u = 0$.  Notice that Eq.~\eqref{eqn:micro} has been obtained
by taking the microscopic limit and is only valid for eigenvalues well below
the chiral scale. For $u=0$ we find that 
$\Delta\lambda_n = 0$ also beyond the microscopic domain and conclude
that in this case the topological domain does not extend beyond the
microscopic domain.  
For $u \ne 0$, however, the situation is
completely different.  All eigenvalues are in the topological domain
and only the first few eigenvalues show universal behavior. 
Comparing the results for $N=400$ and $N=800$ in Fig.~\ref{fig:fig2}, we observe that
the universal domain, i.e., the domain where the eigenvalue ratio 
$\Delta \lambda_n$
does not depend
on $u$, increases with $N$ proportional to $\sqrt N$. This is
in agreement with
microscopic universality for $u < u_c = 1/\Sigma$, which states that 
the distribution of low-lying 
eigenvalues is universal after rescaling them by 
the chiral condensate. 
If we consider the Dirac spectrum around $x$, 
the correction terms to this universal behavior are of the order $Nx^2$.
This implies that the number of eigenvalues
with universal fluctuations around
$\lambda =0$ scales with $\sqrt N$.
 
\begin{figure}[!t]
  \begin{center}
    \includegraphics[width=8cm]{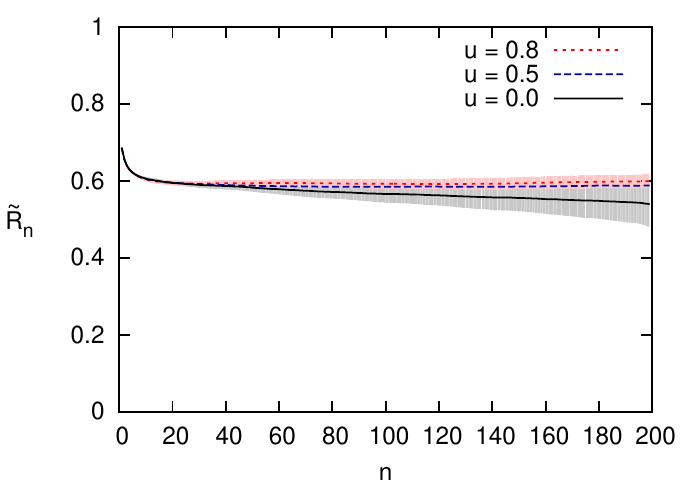}
  \end{center}
  \caption{Convergence of the ratio $\tilde R_n$ of determinants for
    $\nu =2$ and $\nu = 0$ as a function of the number $n$ of
    eigenvalues included.  Results are for an ensemble of $10^6$
    $400\times 400$ matrices.  The shaded areas correspond to the
    statistical errors.}
  \label{fig:fig3} 
\end{figure}

Based on Fig.~\ref{fig:fig2}, a plausible explanation for the
behavior of the ratio of the determinants seen in
Fig.~\ref{fig:fig1} can be given in terms of the $u$ dependence of the
average position of the eigenvalues.  For this reason we plot in
Fig.~\ref{fig:fig3} the same ratios as in Fig.~\ref{fig:fig1}, but
normalized with respect to the average positions of the
eigenvalues. The ratio $\tilde R_n$ defined by
\begin{align}
  \tilde R_n \equiv 
  \frac{\braket{\prod_{k=1}^{n-1} \left(\lambda_k^{\nu=2} /
    \braket{ \lambda_k^{\nu=2} }\right)^2  }}
  {\braket{\prod_{k=1}^{n} \left(\lambda_k^{\nu =0}  /
    \braket{ \lambda_k^{\nu=0}}\right)^2 }}
  \label{eqn:rescale}
\end{align}
is shown for $u =0.0$, $u=0.5$, and $u=0.8$.

We conclude that the $u$ dependence of the ratio of the determinants
is almost exclusively due to the effect of $u$ on the average position
of the eigenvalues.

\begin{figure}[!t]
  \begin{center}
    \includegraphics[width=8cm]{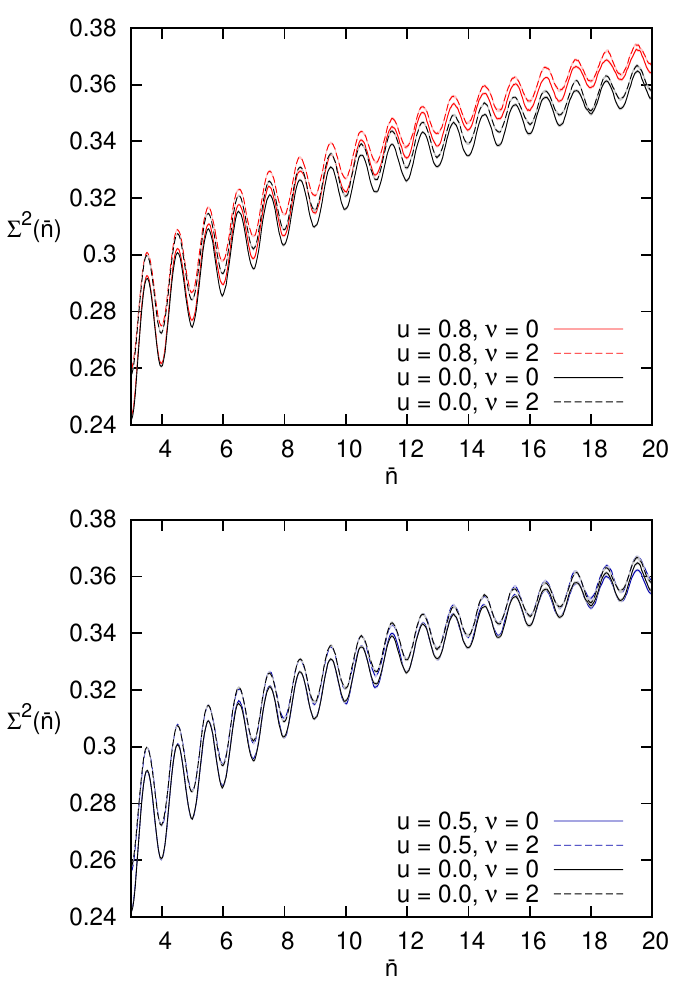}
  \end{center}
  \caption{Number variance versus the average number $\bar n$ of levels in the
    interval $[0, n]$ for an ensemble of $10^6$ $400 \times 400$
    matrices.  The curves for $u=0.0$ and $u=0.5$ only start to
    deviate from each other at $\bar n \ge 15$.}
  \label{fig:fig4} 
\end{figure}

In the theory of disordered systems, a frequently used measure to
test the breakdown of universality is the number variance \cite{Guhr:1997ve}.
This is the variance of the number of levels in an interval
containing $\bar n$ eigenvalues on average. In
Fig.~\ref{fig:fig4} we display the number variance $\Sigma^2$ 
versus the average 
number $\bar n$ of eigenvalues in an interval starting at zero.  
The curves for $u=0.0$ and $u=0.5$ coincide for $\bar n\le 15$, 
while the deviations between $u=0.0$ and $u=0.8$ are already 
significant for $\bar n \ge 5$.  This is in agreement with the 
discussion of Fig.~\ref{fig:fig2}.

\begin{figure}[t!]
  \begin{center}
    \includegraphics[width=8cm]{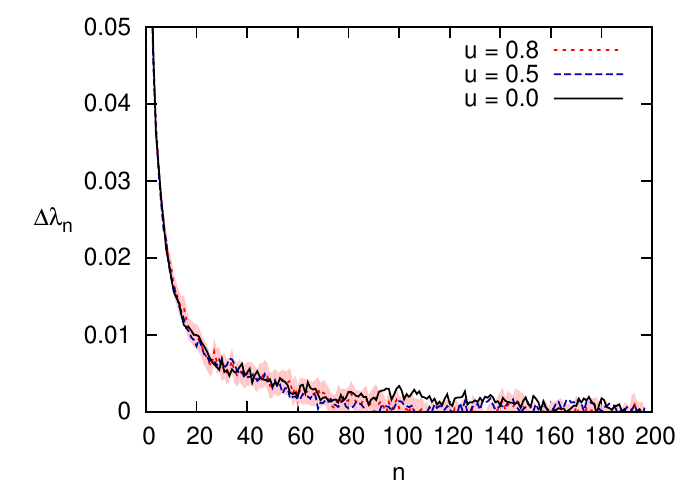}
    \includegraphics[width=8cm]{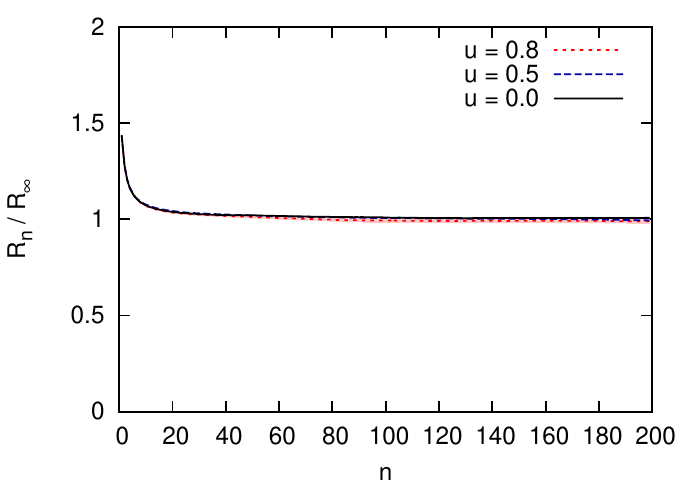}
  \end{center}
  \caption{Topological shift $\Delta \lambda_n$ and the ratio $R_n /
    R_\infty$ for an ensemble of $10^6$ $400 \times 400$ matrices with
    $N_1/N = 0.75$ for model B \cite{Janik:1999ps}.}
  \label{fig:fig5}
\end{figure}

In Fig.~\ref{fig:fig5} we show the behavior of the Dirac eigenvalues
in model B.  We observe that in this model the topological domain does
not extend beyond the microscopic domain even for $u\ne0$.  This is
also the case for model C, which at $u \ne 0$ is equivalent to model A
at $u=0$ after rescaling the chiral condensate $\Sigma\to\Sigma^C(u)$.
The results for model C are therefore identical to the $u=0$ results
in Figs.~\ref{fig:fig1}--\ref{fig:fig4}.  We thus have a further piece
of evidence that nontrivial normalization factors $\N_\nu$ only appear
if the topological domain extends beyond the microscopic domain.

\section{Topological and pseudoscalar susceptibility}
\label{sec:susc}

As mentioned in the introduction, the $\theta$ dependence of the QCD
partition function is obtained by introducing left-handed and
right-handed quark masses according to $z=m e^{i\theta}$ and $z^* = m
e^{-i\theta}$, respectively; see Eq.~\eqref{eqn:theta-gen}.  Denoting
the left-hand side of Eq.~\eqref{eqn:ZQCDtheta} by $Z(z, z^*)$, with the
superscript QCD omitted for simplicity, the topological susceptibility
at arbitrary $\theta$ angle  
is given by
\begin{align}\nn
  \chi_t &= \frac 1V \left(\braket{\nu^2}-\braket{\nu}^2\right) 
  = -\frac 1V \del^2_\theta \log Z(z,z^*)\\
  \label{eqn:chi}
  &= \frac 1V \left( z\del_z + z^*\del_{z^*} \right) \log Z(z,z^*) \\\nn
  &\quad+\frac 1V \left[z^2\del_{z}^2 + {z^*}^2\del_{z^*}^2 
    -2zz^*\del_z\del_{z^*}\right]\log Z(z,z^*)\,.
\end{align}
Because $m\del_m=z\del_z+z^*\del_{z^*}$, the first term on the right-hand side
of this equation is equal to $m\abs{\braket{\bar\psi\psi}}$; see
Eq.~\eqref{eqn:cond}.  The second term on the right-hand side of
Eq.~(\ref{eqn:chi}) is equal to $m^2$ times the pseudoscalar (PS)
susceptibility given by
\begin{align}
m^2\chi_\text{PS}&= V \braket{
(z \bar \psi_L \psi_R -z^*\bar \psi_R\psi_L)^2}_{N_f=1} \nn \\
&\quad-V \braket{ z\bar \psi_L\psi_R - z^* \bar \psi_R \psi_L}_{N_f=1}^2\,.
\label{eqn:ps}
\end{align}
Thus Eq.~\eqref{eqn:chi} becomes 
\begin{align}
  \label{eqn:chit}
  \chi_t=m \abs{\braket{\bar\psi\psi}}+m^2\chi_\text{PS}\,.
\end{align}
This is the well-known chiral Ward identity relating $\chi_t$ to the
chiral condensate and the pseudoscalar susceptibility
\cite{Crewther:1977}.  
Note that $\abs{\braket{\bar\psi\psi}}=\Sigma\cos\theta+O(m)$.

\begin{figure*}[!t]
  \begin{center}
    \subfigure[~$mN\Sigma\gg1$, $\N_\nu = 1$.]{
      \includegraphics[width=8cm]{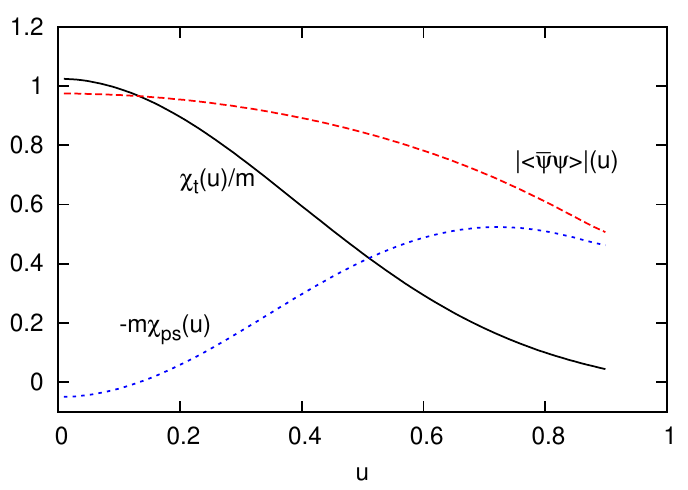}}
    \subfigure[~$mN\Sigma\gg1$, $\N_\nu$ from Eq.~\eqref{eqn:znor}.]{
      \includegraphics[width=8cm]{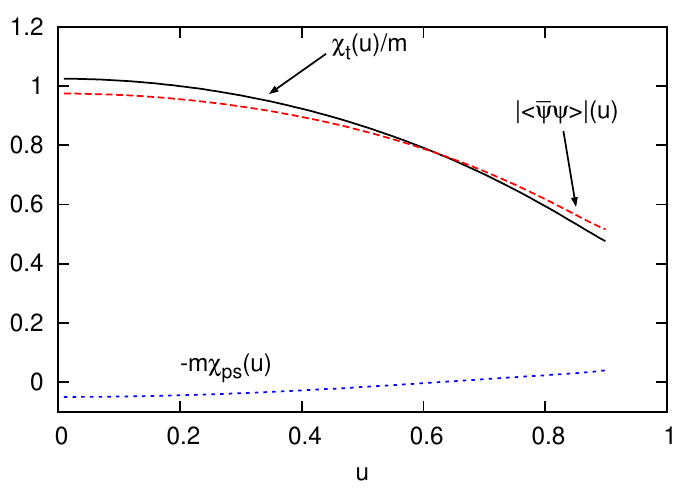}}
    \subfigure[~$mN\Sigma\ll1$, $\N_\nu = 1$.]{
      \includegraphics[width=8cm]{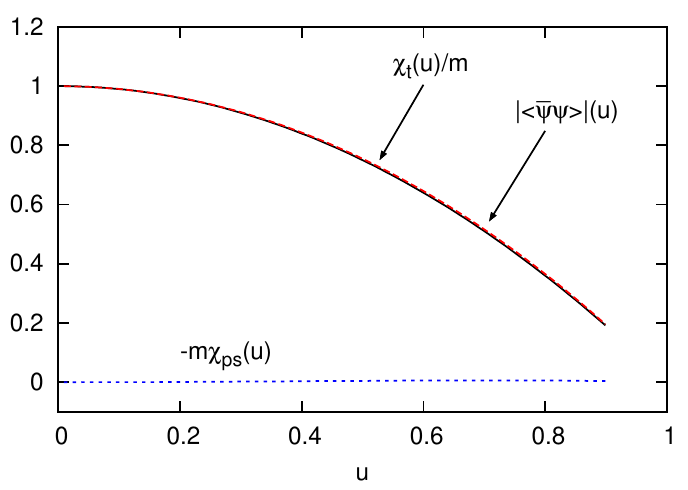}}
    \subfigure[~$mN\Sigma\ll1$, $\N_\nu$ from Eq.~\eqref{eqn:znor}.]{
      \includegraphics[width=8cm]{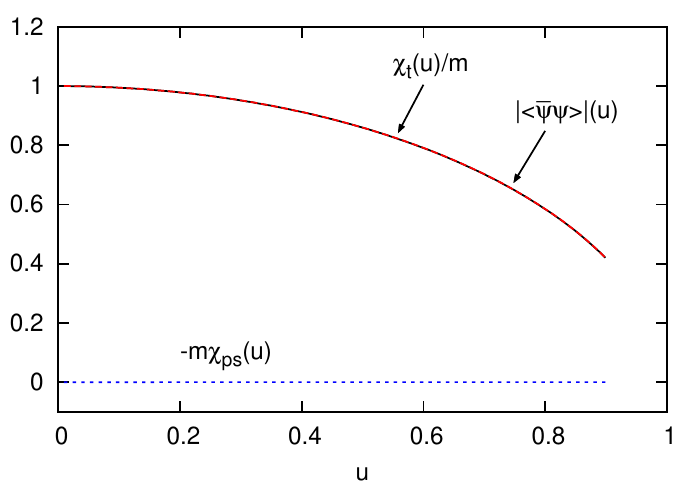}}
    \caption{\label{fig:fig6} Contributions to $\chi_t^A$ for $N=10^3$
      and $m=5\cdot 10^{-2}$ (top) and $m=5\cdot 10^{-5}$ (bottom)
      with or without the normalization factor $\N_\nu$.  We set
      $\theta=0$ and $\Sigma=1$.  The curves were obtained by
      numerical evaluation of Eq.~\eqref{eqn:zthetasum} in connection
      with Eq.~\eqref{eqn:chi}.}
\end{center}
\end{figure*}

The random matrix partition function $Z^A(m,\theta)$ with $m$ in the
ergodic domain can be calculated explicitly from
Eq.~\eqref{eqn:part-theta}, setting $\P_\nu = 1$ according to the
discussion in Sec.~\ref{sec:pnu}.  We will set $\N_\nu =
\tau^{-\abs{\nu}(1-\epsilon)}$, where setting $\epsilon$ to $0$ or
$1$ allows us to switch between including or not including $\N_\nu$.

We first replace the Bessel function $I_\nu$ in Eq.~\eqref{eqn:znurad}
by the integral representation
\begin{align}
  I_\nu(x) = \frac 1{2\pi}\int_0^{2 \pi} d\phi\: e^{i\nu \phi + x \cos
    \phi}\,, 
\end{align}
sum the resulting geometric series in $\nu$, and perform a
saddle-point approximation of the radial integral including
next-to-leading order corrections in $m$ to find
\begin{align}
  Z^A(m,\theta)
  &\sim \int_0^{2\pi} d\phi\: \frac{1-\tau^{2\epsilon}}
  {1-2 \tau^\epsilon \cos\phi + \tau^{2\epsilon}} \nn\\
  &\quad\times \exp\left[m N \Sigma \tau \cos(\phi-\theta)\right]\nn\\
  &\quad\times
  \exp\left[\frac1{4\tau^2} N m^2 \Sigma^2\cos^2(\phi-\theta)\right].
  \label{eqn:zthetasum}
\end{align}
Note that
\begin{align}
  \lim_{u \to 0} \frac{1-\tau^{2\epsilon}}{2\pi(1-
    2\tau^{\epsilon} \cos\phi+\tau^{2\epsilon})} &= \delta(\phi)
  \intertext{but also}
  \lim_{\epsilon \to 0} \frac{1-\tau^{2\epsilon}}{2\pi(1-
    2\tau^{\epsilon} \cos\phi +\tau^{2\epsilon})} &= \delta(\phi)\,.
\end{align}
Therefore for $\epsilon \to 0$ or $u \to 0$ we find
\begin{align}
  Z^A(m,\theta) \sim \exp\left[m N \Sigma \tau \cos\theta+
    \frac{1}{4\tau^2}Nm^2 \Sigma^2\cos^2\theta\right]
\end{align}
and thus by Eq.~\eqref{eqn:chi}
\begin{align}
  \chi_t^A(u)& = m \Sigma^A(u) \cos\theta + O(m^2)\,,
  \label{eqn:chistandard}
\end{align}
which is consistent with results obtained by Crewther
\cite{Crewther:1977}.  We conclude that for $u=0$ or if we include
the normalization factor \eqref{eqn:znor} the contribution of the
pseudoscalar susceptibility vanishes in the chiral limit.

The situation is different, however, if we do not include the
$\N_\nu$. 
For $m N \Sigma \gg 1$ the
contribution of the pseudoscalar susceptibility to the topological
susceptibility becomes comparable to that of the chiral condensate but
with opposite sign and thus leads to a significant suppression of the
topological susceptibility (see Fig.~\ref{fig:fig6}). Because the 
saddle-point approximation breaks down
close to $u=1$ we do not plot the curves
of Fig.~\ref{fig:fig6} in this region.
For $mN\Sigma \ll 1$ the exponent  in Eq.~(\ref{eqn:zthetasum}) can
be expanded, and after evaluating the integral 
analytically we find
\begin{align}
Z^A(m,\theta) \sim 1+mN\Sigma \tau^{1+\epsilon} \cos \theta\,.
\end{align}
This result agrees with
Fig.~\ref{fig:fig6} and shows that in this limit the
contribution of the pseudoscalar susceptibility at $u\ne 0$ is small
also without $\N_\nu$.  

Metlitski and Zhitnitsky have recently
found another situation in which the $O(m^2)$ term in
Eq.~\eqref{eqn:chit} becomes important, i.e., the superfluid
phase of QCD with two or three colors
\cite{Metlitski:2005}.  Of course, if we include
the $\N_\nu$ in model A (as we should) we do not see this effect.
Nevertheless, our observation may potentially be of importance; see
the conclusions.

For models B and C no normalization factors $\N_\nu$ are needed 
to ensure a vanishing contribution of the pseudoscalar susceptibility.

The vanishing of the contribution of the pseudoscalar susceptibility
also imposes constraints on the $\nu$-dependence of pseudoscalar
correlators and can be used as a check of results that were recently
derived for the $\epsilon$ domain
\cite{Akemann:2008vp,Bernardoni:2008ei}.

\section{Conclusions}
\label{sec:conc}

It is well-known that random matrix models for QCD at {\it zero}
imaginary chemical potential (or temperature) $u$ have 
the correct $\theta$ dependence.  In this paper we have shown that
this is not automatically the case for $u \ne 0$. 
We obtain the correct $\theta$ dependence only after introducing
$\nu$-dependent normalization factors $\N_\nu$ in the
sum over topologies.

To explain this we have introduced the topological domain of the Dirac
spectrum, which is defined as the part of the Dirac spectrum that is
sensitive to the topological charge.  We have shown that for $u = 0$
the topological domain coincides with the microscopic domain.
This is also the case at $u\ne0$ for models for which no
$\nu$-dependent normalization factors are needed to obtain the correct
$\theta$ dependence.  However, for the model we analyzed that requires
nontrivial normalization factors, the complete 
Dirac spectrum is inside the topological domain.  This results in a
partition function that gives universal behavior for small Dirac
eigenvalues, but has bulk spectral correlations that depend both on
$u$ and on the topological charge.  In the thermodynamic limit this
leads to an additional $u$-dependent factor in the partition function
at fixed topological charge which results in an incorrect
$\theta$ dependence of the partition function. To obtain a partition
function with the usual behavior in the chiral limit, one has to
introduce additional $\nu$-dependent normalization factors in the sum
over topologies.

Our observations are of potential importance for lattice QCD at
nonzero imaginary chemical potential or temperature.  Depending on,
e.g., the fermion formulation or the algorithm used, it could be that
nontrivial normalization factors are needed in the sum over
topological sectors, and these could even persist in the continuum
limit.   To find out whether such normalization factors might
be necessary, it would be interesting to determine the topological
domain as a function of the deformation parameters.  This is feasible
with
current lattice technology.  To be consistent
with the general properties of QCD, the topological domain
should not extend beyond the microscopic domain.
Future work will tell us if this interesting picture prevails.

{\bf Acknowledgments.} We thank T.~Hatsuda and J.~Osborn for helpful
discussions and acknowledge support by BayEFG (C.L.), the Humboldt
Foundation (M.O., J.V.), U.S.~DOE Grant No.~DE-FAG-88ER40388 (J.V.), DFG and
JSPS (T.W.).  T.W.\ thanks the Theoretical Hadron Physics Group at Tokyo
University for their hospitality.

\end{document}